# The Confirmation of Three Faint Variable Stars and the Observation of Eleven Others in the Vicinity of Kepler-8b by the Lookout Observatory

Neil Thomas and Margaret Paczkowski
Department of Astronautical Engineering, United States Air Force Academy, CO 80840, USA; neil.thomas@afacademy.af.edu

**Abstract**

A commissioning survey of the Lookout Observatory has observed fourteen faint ($V \sim 13$ to $17$) variables in the region of the exoplanet Kepler-8b. Three of these are variable star candidates discovered by the Asteroid Terrestrial-Impact Last Alert System (ATLAS) and confirmed here. The ATLAS survey identified 315,000 probable variables within its wide-field survey in 2018. The faintness (down to $r \sim 18$) and small amplitudes (down to 0.02 mag) included in these candidates makes external validation difficult. The confirmation of this handful of variable stars lends credibility to the ATLAS catalog. Lastly, the agreement between various surveys and this new one validates the use of this instrument for variable star and exoplanet research.

## 1. Introduction

The primary function of the Lookout Observatory (LO) is to observe exoplanets via the transit photometry method. It routinely attains photometric precision better than 0.002 mag (2 mmag) for bright targets. It autonomously performs photometry on all suitable stars within the field of view (FOV) in addition to the targeted exoplanet host star. It thus provides valuable observations of many variable stars during a given night.

Section 2 discusses the characteristics and expected performance of the LO in contrast to several other sources of photometry. Section 3 validates results against external data, provides light curves for 14 variable stars, and confirms the existence of three variable star candidates identified by ATLAS. Section 4 is the conclusion.

## 2. Instrumentation and methods

The LO is optimized to maintain the photometric precision necessary to observe exoplanet transits. Even large, short-period exoplanets generally only cause a stellar dimming of 30 mmag or less and have a duration of two to three hours. To decisively capture a transit (and a sufficient baseline before and afterwards), noise levels of less than 2 mmag are routinely maintained from dusk to dawn for 10$^{th}$ magnitude stars using one-minute exposures. However, the LO generally collects useful photometry down to *V*<17, depending on conditions and exposure length. A single FOV is observed with dimensions: 114' x 86' or 2.7 deg$^2$. This allows the LO to collect photometry on 500 to 15,000 stars with a photometric cadence of 10 to 60 seconds. Optical filters are not usually used. Within the past two years the LO has successfully observed approximately 70 exoplanet transits. The results presented here are from a single night. In summary, LO provides high-cadence and nearly continuous photometry for up to 15,000 stars over the time span of a night. Appendix 1 provides more detailed performance characteristics.

Images are collected semi-autonomously using MAXIM DL and CCDCOMMANDER. A custom-made software pipeline developed in MATLAB performs the photometry. The software is specifically designed to mitigate the impact of observing from a light-polluted site in Colorado which also typically experiences variable cloud cover each night. The site has a light pollution Bortle index of four, with sky brightness being measured at 20 mag per square arc second on moonless nights.

This paper uses photometry from several other external surveys to validate ours. The ATLAS survey is a wide-field survey with the primary purpose of detecting hazardous asteroids approaching Earth (Tonry et al. 2018). It has incidentally become a rich source of detections for other transient evens such as supernovae, flares, gamma-ray burst, and variable stars down to *r*~19. Recently, a dedicated list of new variable star candidates was released (Heinze et al. 2018). In contrast to the high cadence of observations collected during transit photometry, ATLAS generally makes a few photometric measurements of any given star each night. ATLAS identifies variables and catalogs them using complex frequency analysis techniques and many months of such observations. While the ATLAS archive of new variables is significant, the faintness and low amplitude of many of these variables makes verification difficult. Yet LO also identified 14 of the variables detected by ATLAS. Three of them were only previously identified by ATLAS as candidates and thus likely do not appear in the American Association of Variable Star Observers (AAVSO) catalog for this reason.

The All-Sky Automated Survey for Supernovae (ASAS-SN) is an all-sky survey which collects photometry on stars down to *V*<17 every two to three days in the search for supernovae (Jayasinghe et al. 2018). The survey has incidentally identified over 66,000 variable stars and made their light curves available to the public. Although the observations are often spaced days apart, short-period variables are detectable thanks to an archive of observations extending back to 2014 and frequency analysis techniques.

Unlike ATLAS and ASAS-SN, the mission of the SuperWASP project is exoplanet discovery, and as a result, it has been responsible for the discovery of nearly 200 exoplanets (Pollacco et al. 2006). It can survey the entire sky every 40 minutes, but at a modest accuracy of 1% (or ~ 11 mmag) for bright targets (*V*=7-11.5). This level of accuracy is sufficient to detect large exoplanets, particularly when its long baseline of observations allows for phase folding many periods to reduce noise.

The Kepler spacecraft collected photometry with the precision necessary to detect the transits of Earth-sized planets (Borucki et al. 2010). Its primary mission monitored a relatively small FOV in the region of Cygnus and Lyra. As a result, the quality of the nearly continuous photometry of stars within this region is unprecedented.

The LO observed a FOV centered on Kepler-8b (RA 281.2881°, DEC 42.4511° J2000) on 7 June 2020, and the number of stars ultimately providing useful light curves was 4,508. The data pipeline flagged merely 14 stars (0.35%) as variables. This low fraction is largely attributed to the fact that LO detections are limited to variables having a period comparable to or less than the 5.3-hour observation window. Three of these variables were not identified as variables in any other survey source except as ATLAS candidates. Existing characteristics for these 14 stars are provided in Table 1. Targets 4, 10, and 14 did not appear in the AAVSO catalog of variable stars, likely because ATLAS has been the only source to report them as candidates.

Table 1. Existing characteristics of the 14 variable stars observed by LO. ID # 4, 10, and 14 are ATLAS candidates confirmed in this paper.

| ID | Name[1] | RA[2] | DEC[2] | Mag[1] | Period[1](days) | Range[1] | Type[1] | Discoverer |
|---|---|---|---|---|---|---|---|---|
| 1 | ROTSE1 J184234.00+420947.9 | 280.6414 | 42.1635 | 13.440 | 0.319774 | 0.403 | RRC | ROTSE |
| 2 | ROTSE1 J184517.00+424010.4 | 281.3238 | 42.6700 | 13.380 | 0.807725 | 0.277 | ELL | ROTSE |
| 3 | ASASSN-V J184116.40+421342 | 280.3183 | 42.2284 | 13.430 | 0.175226 | 0.050 | DSCT | ASAS-SN |
| 4 | 2MASS 18441165+4201591 | 281.0486 | 42.0331 | | | | | ATLAS candidate(dub) |
| 5 | KIC 7176440 | 282.5124 | 42.7709 | 14.293 | 0.358267 | 0.115 | ECL | Kepler |
| 6 | KIC 7173910 | 281.2031 | 42.7461 | 14.360 | 0.402244 | 0.132 | EW | Kepler |
| 7 | WISE J184227.5+422724 | 280.6149 | 42.4568 | 12.429 | 0.846410 | 0.357 | EA | WISE |
| 8 | KIC 6836820 | 281.4374 | 42.3286 | 14.500 | 0.227270 | | DSCT | Kepler |
| 9 | KIC 6836140 | 281.0407 | 42.3978 | 14.646 | 0.487721 | 0.245 | SD (EW) | Kepler |
| 10 | 2MASS 18452610+4231055 | 281.3588 | 42.5181 | | | | | ATLAS candidate (EW) |
| 11 | MarSEC_V13 | 280.2693 | 42.6732 | 15.65 | 0.333045 | 0.400 | EW | Mar-SEC |
| 12 | V0351 Lyr | 282.3584 | 42.9808 | 15.25 | 0.839481 | 1.100 | AHB1 | C. Hoffmeister (1966) |
| 13 | CSS_J184816.3+414748 | 282.0681 | 41.7965 | 16.04 | 0.600642 | 1.190 | RRAG | Catalina |
| 14 | 2MASS 18465788+4156020 | 281.7412 | 41.9339 | | | | | ATLAS candidate (EW) |

[1] typically from the discovery source (magnitudes zero points can vary greatly based on the survey wavelength sensitivity)
[2] from GAIA, J2000

## 3. Results

This section seeks to validate the LO survey by comparing these results to two variables in this FOV having available photometry from several other surveys. Light curves for all 14 variables are subsequently shown, including the three ATLAS candidates that this survey confirms.

3.1 External Validation

The availability of external photometry for these variables is given in Table 2. To demonstrate the validity in this survey, LO results are first compared for a known variable that has been observed by all four surveys. The star KIC 7173910 is a W Ursae Majoris-type eclipsing variable (EW) identified initially in Kepler data (Prsa et al. 2011) and is ID # 6 in this survey. It is *V*=14.36 and has a period of 0.402247 days (Prsa et al. 2011) with an amplitude of 264 mmag. ATLAS observed this variable 151 times over several years and correctly classified it as a contact or near-contact eclipsing binary. Photometry obtained from CasJobs (see acknowledgments) was phase folded to the known period. Kepler and SuperWASP data were then retrieved from the NASA Exoplanet Archive (see acknowledgments). ASAS-SN photometry was obtained directly from the survey's website (Jayasinghe et al. 2018). The phase folded measurements from all these surveys are shown in Fig. 1. The detections are clear and consistent among all surveys.

Table 2. Light Curve availability[1].

| ID | Kepler[2] | SuperWASP[2] | ASAS-SN[3] | ATLAS[4] | Notes |
|---|---|---|---|---|---|
| 1 | y | y | y | | |
| 2 | y | y | y | y | |
| 3 | | | y | y | |
| 4 | | | | y | Only identified by ATLAS |
| 5 | y | y | | y | |
| 6 | y | y | y | y | First test case, Figs. 1 & 2 |
| 7 | | | | y | Fig 6. STOCH detection by ATLAS of a known EA. |
| 8 | y | y | y | y | |
| 9 | y | y | y | y | Second test case, Figs. 3 & 4. |
| 10 | | | y | y | Only identified by ATLAS |
| 11 | | | y | y | |
| 12 | | | y | y | |
| 13 | | | y | y | |
| 14 | | | y | y | Only identified by ATLAS |

[1] Availability is only marked when the survey also identified a star as variable. Photometry may actually exist for these stars.
[2] https://exoplanetarchive.ipac.caltech.edu/
[3] https://asas-sn.osu.edu/variables
[4] http://mastweb.stsci.edu/

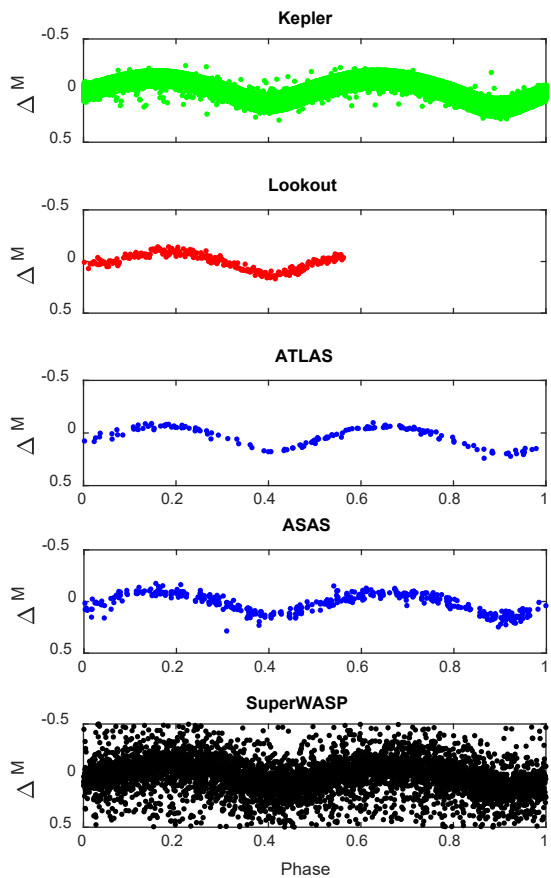

Figure 1. Phase folded light curves for the EW variable KIC 7173910 (ID #6) from multiple surveys. This star is *V*=14.36 and has a period of 0.402244 days. LO results compare well with ATLAS and ASAS-SN. Kepler and SuperWASP light curves appear deceptively noisy. But in fact, the much larger sizes of their datasets allows for greatly improved results once binned.

To ensure equity among the comparisons, each survey is binned to have the same density of data in phase space that appears in the sparsest survey, ATLAS in this case. Binning reduces the number of data points but also reduces their statistical scatter so long as the noise is Gaussian (white) as opposed to that caused by instrumental systematics (red). Kepler provides nearly 200,000 data points and is effectively noise-free once binned. Figure 2 shows these binned light curves. ASAS-SN provides 210 raw data points prior to binning, and the SuperWASP data includes 8,573 observations. The value of SuperWASP's persistence in collecting many medium-precision observations is the most evident after binning. The LO results are favorably comparable to ATLAS and ASAS-SN in quality even though LO photometry comes from a single night. Unfortunately, the LO observation duration did not allow for coverage of a full period.

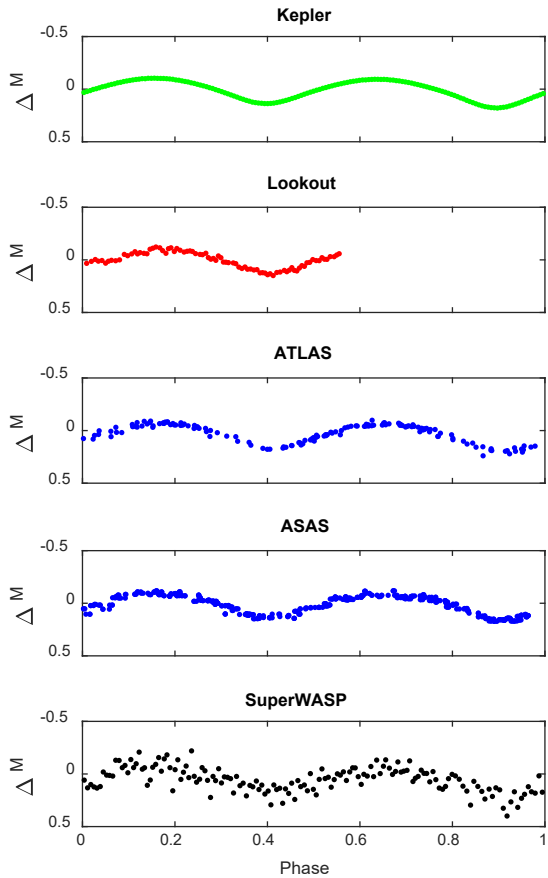

Figure 2. Phase folded light curves for KIC 7173910 (ID #6) with data binned to a common sampling to fairly compare surveys having widely varying numbers of observations. LO results compare well with other ground-based surveys. SuperWASP noise levels are greatly reduced, and Kepler data appears free of noise after binning.

The same approach is followed for the similar study of a star having the same sources of independent data. The star KIC 6836140 (ID # 9) has a magnitude of *V*=14.646 and is classified by Kepler as a semi-detached (SD) eclipsing binary, a subclass of EW, with a period of 0.487721 days (Prsa et al. 2011). ATLAS and ASAS-SN both classify it as an EW with an amplitude of 270 mmag.

Figure 3 shows the binned light curves of this star from various surveys. All sources agree, including LO results.

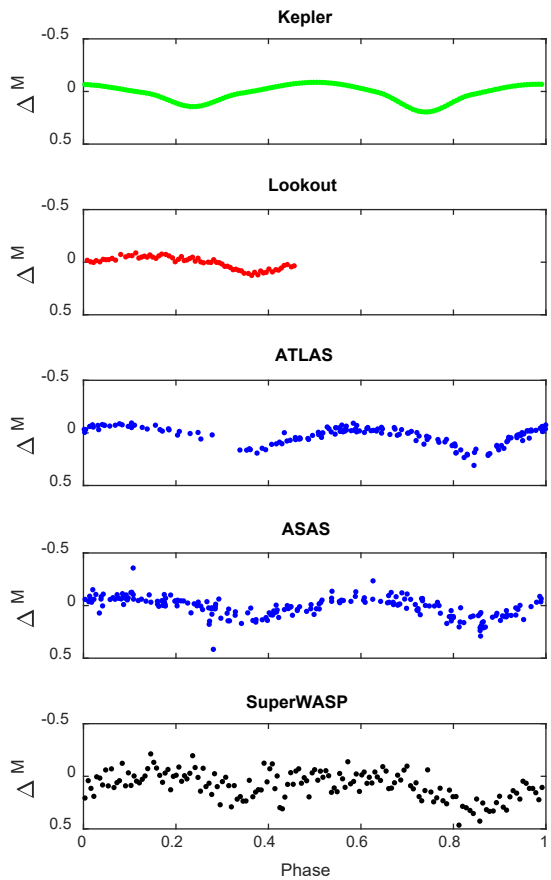

Figure 3. Phase folded light curves for the EW KIC 6836140 (ID # 9). This star is $V$=14.646 and has a period of 0.487721 days. LO results compare well with other ground-based surveys.

A comparison between non-binned LO and binned Kepler results is shown in Fig. 4. If Kepler data is considered to be virtually noise free in this case and the two are subtracted, then we have a nearly direct measure of LO noise levels. Doing so yields an RMS of 28.4 mmag, which is consistent with our design goals for a star of this magnitude (see Appendix 2).

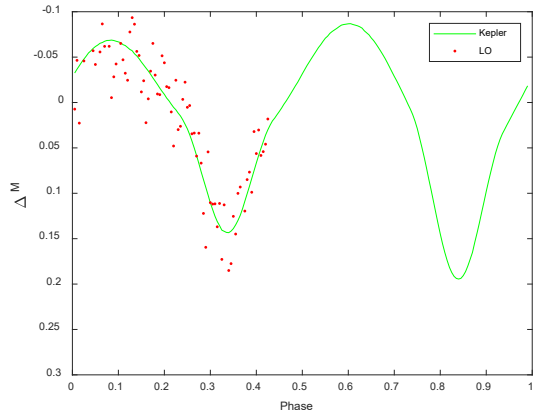

Figure 4. Non-binned LO data (dots) compared to binned Kepler data (solid). Kepler results are practically noise free at this scale and the deviation in LO data from Kepler can be used to measure our noise at 28.4 mmag for this *V*=14.6 star.

3.2 Photometric Results

Having confidence in the LO light curves, Fig 5 presents the light curves and results for all 14 variable stars detected in this survey. Table 3 provides characteristics derived from this work, when possible.

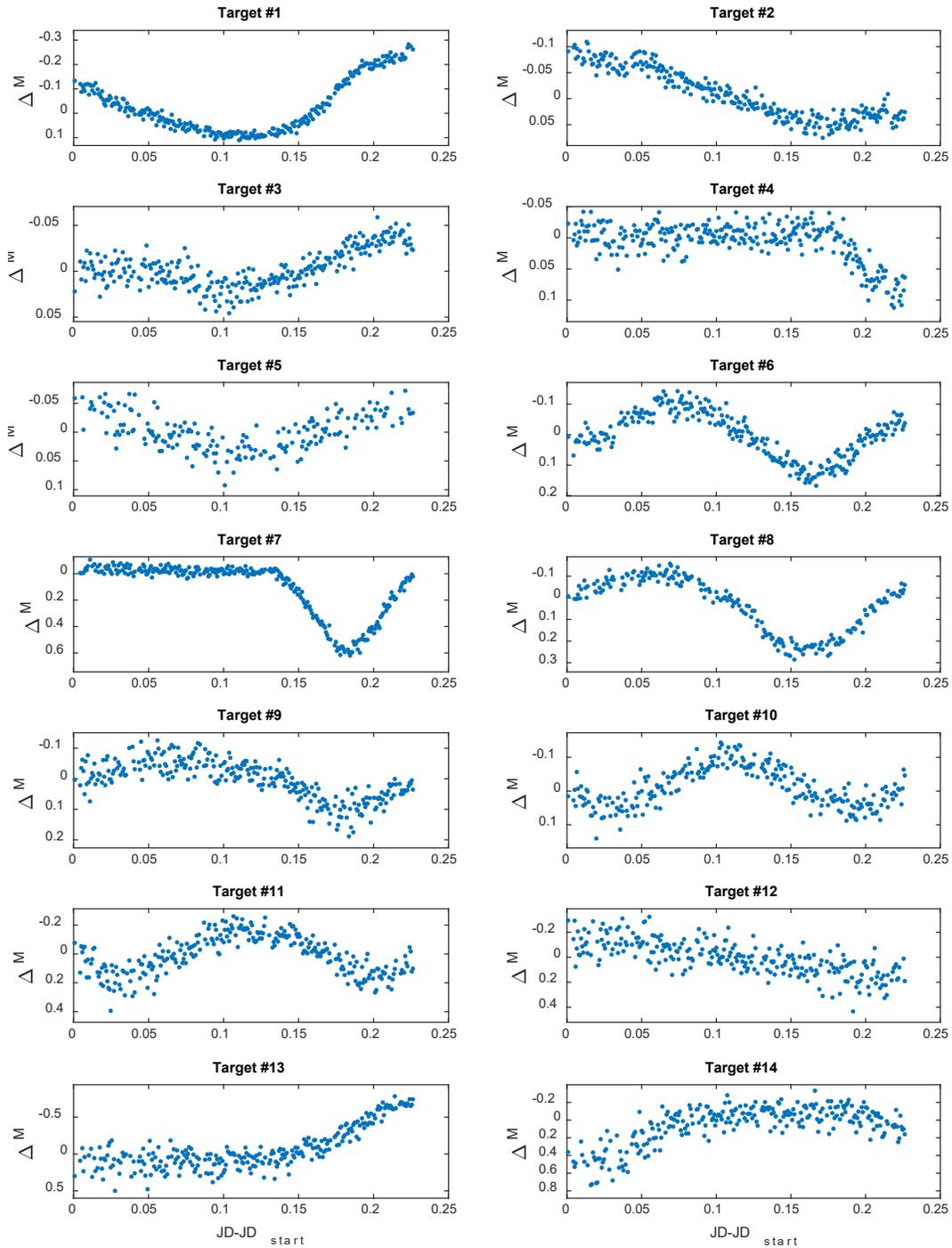

Figure 5. The LO light curves of the 14 variables automatically detected by our software. Targets 4, 10, and 14 do not appear in the AAVSO catalog of variable stars, likely because they have only been previously reported as ATLAS candidates.

Table 3. Characteristic derived by Lookout observations.

| ID | Mag[1] | Period[2] | Range[2] | Notes |
|---|---|---|---|---|
| 1 | 13.1628 | | | |
| 2 | 13.0281 | | | |
| 3 | 13.325 | | | |
| 4 | 14.15 | | | |
| 5 | 14.2977 | 0.44 | 0.072 | |
| 6 | 14.3498 | 0.36 | 0.23 | Figure 2. First validation star. |
| 7 | 14.4245 | | 0.6 | Figure 6. |
| 8 | 14.3605 | | 0.35 | |
| 9 | 14.747 | 0.46 | 0.147 | Figure 3. Second validations star. |
| 10 | 14.7456 | 0.32 | 0.142 | |
| 11 | 15.5203 | 0.34 | 0.314 | |
| 12 | 15.9147 | | | |
| 13 | 16.1783 | | | |
| 14 | 16.0793 | | | |

[1] as derived from calibrating all observed stars to Gaia magnitudes
[2] only provided for sinusoidal targets where a significant portion of the period was observed. Should not be taken as superior to any existing values.

### 3.3 ATLAS Candidate Confirmations

This survey has confirmed three candidate variables previously only identified by ATLAS. They range in magnitude from *V*=14.15 to 16.08. ID #4 (2MASS 18441165+4201591) is an ATLAS candidate classified as dubious by their survey. This survey, however, confirms a significant variability. The observation window range is too limited to characterize it completely, but it seems to be a transient event such as an EA eclipsing binary. ID # 10 (2MASS 18452610+4231055) clearly shows sinusoid variability, in agreement with the ATLAS classification as an EW. The results for ID #14 (2MASS 18465788+4156020) do not show a significant portion of the period but the LO light curve is consistent with the ATLAS identification as an EW.

We also match a star classified by ATLAS as STOCH to the known EA WISE J184227.5+422724 (Chen et al. 2018). The STOCH classification implies "variables that do not fit into any coherent periodic class" (Heinze et al. 2018). The inability of ATLAS to classify this variable (ID #7 in our survey) is likely due to the inherent difficulty in the detection of transient events using tradition frequency analysis, which favors sinusoidal patterns (Kovacs et al. 2002). Figure 6 shows the light curves from several surveys. The WISE detection clearly shows the primary and secondary eclipses. The ATLAS photometry demonstrates the eclipsing nature of this system once phase folded to the period established by WISE. For this variable, ASAS-SN observations were poorly sampled during the phase of the transit. Their results are consistent with an EA but would not have been expected to lead to a detection due to their sampling. The LO results clearly show a portion of the primary eclipse.

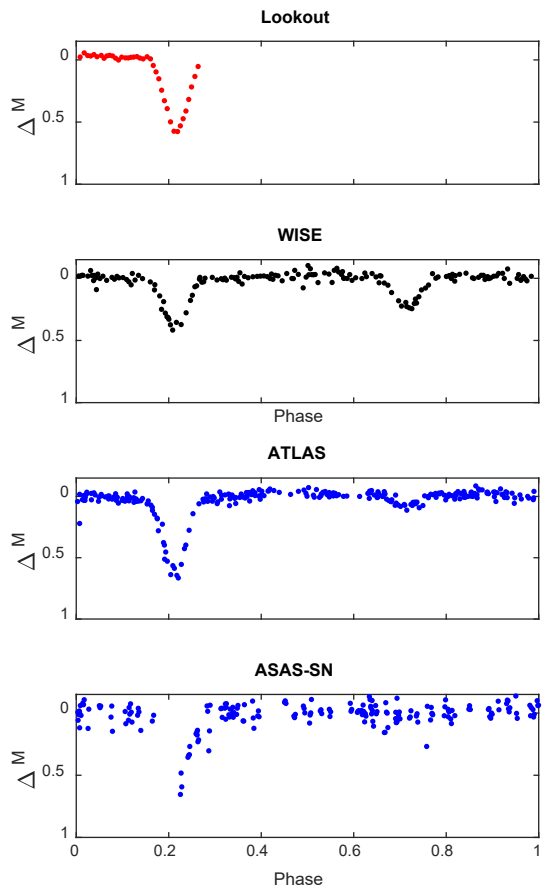

Figure 6. The light curves of variable star ID #7, an EA type binary discovered by WISE. ATLAS and ASAS-SN photometry agrees, although those surveys did not initially classify its type.

### 4. Conclusion

The Lookout Observatory has finished its commissioning phase and demonstrated its photometric goals by automatically identifying short-period variables at a quality that is comparable to other ground-based programs. The data was collected in a single night rather than the years required by the low-cadence surveys (ATLAS and ASAS-SN), for which this is not their primary goal. Of the 14 stars examined here, three ATLAS candidates were confirmed. ATLAS, ASAS-SN, and LO also successfully observed an EA discovered by WISE. The LO has observed many FOVs over the past two years, often multiple times. This survey will continue to focus on exoplanet observation, but variable star data will be released as it becomes available.

**Acknowledgements**

In addition to the support of the Astronautical Engineering Department at the US Air Force Academy, the authors would like to thank the LO construction team, Savannah Jane and P.P., as well as Kyle Ziegler who was instrumental during the proofing process.

CasJobs was used in retrieving ATLAS light curves as authored by the JHU/SDSS team. http://casjobs.sdss.org/CasJobs.

This research has made use of the NASA Exoplanet Archive, which is operated by the California Institute of Technology, under contract with the National Aeronautics and Space Administration under the Exoplanet Exploration Program.

This paper makes use of data from the first public release of the WASP dataset provided by the WASP consortium and services at the NASA Exoplanet Archive, which is operated by the California Institute of Technology, under contract with the National Aeronautics and Space Administration under the Exoplanet Exploration Program.

We acknowledge with thanks the variable star observations from the *AAVSO International Database* contributed by observers worldwide and used in this research.

PA#: USAFA-DF-2020-337

**Appendix 1**

The Lookout Observatory consist of an 11" Celestron telescope modified to f/1.9 with a HyperStar. Imaging is done with a ZWO ASI 1600 CMOS camera. LO was constructed by faculty and students of the Astronautical Engineering Department at the US Air Force Academy and focused on consumer grade instruments (entire observatory less than $10,000 USD). It first successfully observed an exoplanet transit in May of 2019. Although it is at an altitude of 7000', it suffers from suburban light pollution and variable weather. About 20% of nights have long enough periods of clarity to allow for observations.

This team developed a custom data pipeline to allow flexibility in dealing with poor sky conditions as well as issues common to small observatories such as errors in tracking, focus, and mirror flop. It also allows for automated processing of all reasonable stars without user input for targeting, calibration star choice, or photometric aperture selections. Light curves lasting 5.4 hours were produced for 4,508 stars ranging 12-17th magnitude, at a 60 second cadence. The standard deviation of the photometry for each star observed this night is plotted against magnitude in Fig. A.1. Since most stars are stable, the data creates a clear function of expected performance versus magnitude. Photon shot noise is also calculated using the flux of each star and the excess observed noise is used to model scintillation noise and red noise. The standard deviation in magnitude for stable stars of 13th magnitude is approximately 10 mmag and is primarily photon limited. The red noise was calculated to be 0.61 mmag and indicates the ultimate precision possible of any star in this survey, regardless of brightness.

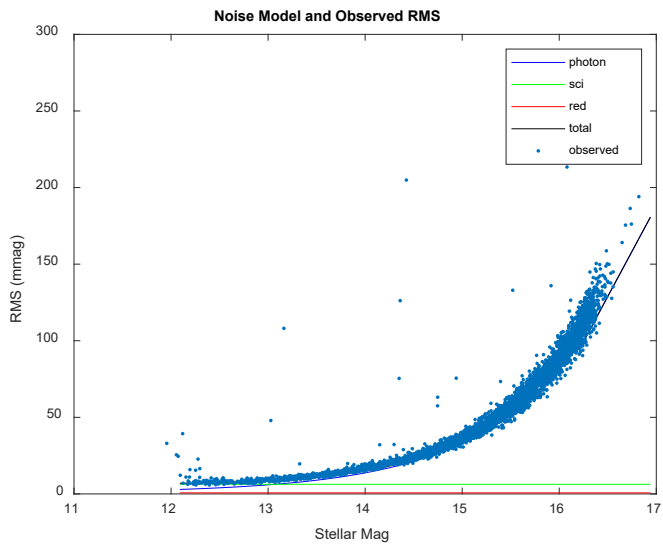

Figure A.1. The performance of the LO survey on this night. The precision of the data degrades as expected with increasing magnitude. The standard deviation of the photometry for each of these mostly stable stars creates a curve very much in accordance with the calculated photon limits. Quality is slightly worse because of atmospheric scintillation and the red noise caused by imperfect instrumentation and programming. The instrumental noise is below 10 mmag for $V<13$.

## Appendix 2

Our primary design goal is maintaining noise levels of less than 2 mmag from dusk to dawn for 10[th] magnitude stars and one-minute exposures. Most stars in this survey are fainter than this. So, we must have a way of scaling results for fainter stars to judge performance.

The photon limited shot noise, σ is the square root of the measured flux, F as seen in Eq. 1, where both are measured in photons.

$$\sigma = \sqrt{F} \tag{1}$$

A brighter star will have greater flux and while its noise will increase, the fraction of noise in the overall signal will decrease. The fraction of the signal that is noise will be given by Eq. 2.

$$\text{noise fraction} = \frac{\sqrt{F}}{F} = \frac{1}{\sqrt{F}} \tag{2}$$

If we compare two stars of different brightness' we can estimate the ratio of precisions with Eq. 3. If flux is increased four times, then the relative error will be cut in half.

$$\frac{\sigma_1}{\sigma_2} = \sqrt{\frac{F_2}{F_1}} \tag{3}$$

The ratio of fluxes between a notional $V=10$ star, $F_{10}$ can be related to the flux of target star, $F_T$ of known magnitude by Eq. 4.

$$\frac{F_{10}}{F_T} = 10^{-0.4(M_{10}-M_T)} \tag{4}$$

We have been using the logarithmic magnitude scale to measure our noise, not flux. While the same relationship between brightness and relative noise does not strictly apply when speaking in magnitudes, it is very close so long as the flux values are large, which is the case here. When we combine Eq. 3 and 4, we can estimate the RMS (in units of magnitude) for a $V=10$ star based on observed error in a target star using Eq. 5. This equation also includes the fact that flux is linearly dependent on exposure time, t.

$$RMS_{10} = RMS_T \sqrt{\left(\frac{t}{60}\right) 10^{0.4(M_{10}-M_T)}} \tag{5}$$

For the example in Sec. 3.1, we have a $V=14.75$ star demonstrating an RMS of 28.4 mmag using 60 second exposures. Eq. 5 yields an equivalent RMS for a V=10 star under these same conditions of 3.18 mmag.

While this performance is slightly poorer than the design objective, it is comparable. Also, this is an analysis of only one star and not a statistical study of all stars in the FOV. A more comprehensive approach is to note in Fig. A.1 that the body of data for all stars indicates a noise level of approximately 10.3 mmag at $V$=13. Using Eq. 5 with this relationship yields an estimated error at $V$=10 of 2.58 mmag. Considering that this is based on a single and unremarkable night, we consider this a successful demonstration of our design goals.